\documentclass{PoS}
\usepackage{graphicx}

\title{Exploring the QCD phase structure with density fluctuations}

\ShortTitle{Exploring the QCD phase structure with density fluctuations}

\author{Chihiro Sasaki\\
Physik-Department,
Technische Universit\"{a}t M\"{u}nchen
D-85747 Garching, Germany\\
E-mail: \email{csasaki@ph.tum.de}}

\author{Bengt Friman\\
GSI, D-64291 Darmstadt, Germany\\
E-mail: \email{b.friman@gsi.de}}

\author{Krzysztof Redlich\\
Institute of Theoretical Physics,
University of Wroclaw,
PL--50204 Wroc\l aw, Poland\\
E-mail: \email{redlich@ift.uni.wroc.pl}}

\abstract{
We briefly summarize the properties of conserved charge
fluctuations as a sensitive probe for the QCD phase transitions.
We discuss the denisty fluctuations which play a significant role
to search for the critical end point.
The importance of spinodal instabilities to distinguish the
first-order phase transition is also indicated.
}

\FullConference{Critical Point and Onset of Deconfinement
          4th International Workshop\\
         July 9-13 2007\\
         GSI Darmstadt,Germany}

\begin{document}

\section{Introduction}

The existence of a critical end point (CEP) in QCD phase diagram is one of the striking
expectations~\cite{cep,srs}, which has been explored based on calculations in effective
models and on universality arguments. The appearance of the CEP in the
temperature $T$ and quark chemical potential $\mu_q$ plane was also studied  in terms of
Lattice Gauge Theory (LGT)~\cite{LGT}.
 The search for the CEP has recently
attracted considerable attention in the context of heavy ion phenomenology~\cite{srs}. 
Thus, it is of particular interest to identify the position of the CEP
in the phase diagram and to study generic properties of thermodynamic quantities in its
vicinity. The analysis of fluctuations is a powerful method for characterizing the
thermodynamic properties of a system. Modifications in the magnitude of fluctuations or
the corresponding susceptibilities have been suggested as a possible signal for
deconfinement and chiral symmetry restoration~\cite{kunihiro,srs,br,qsus:model}. In this
context, fluctuations related to conserved charges are of particular
interest~\cite{fluct}.

In this contribution we briefly summarize the properties of the conserved charge
fluctuations to probe the QCD phase structure. The role of the fluctuations in order to
identify the location of the CEP as well as the phase boundaries is discussed. We also
show  that the enhanced baron or electric charge density fluctuations could signal
the first order phase transition in the presence of spinodal decomposition.

\section{Fluctuations of conserved charges}

\begin{figure}
\begin{center}
\includegraphics[width=8cm]{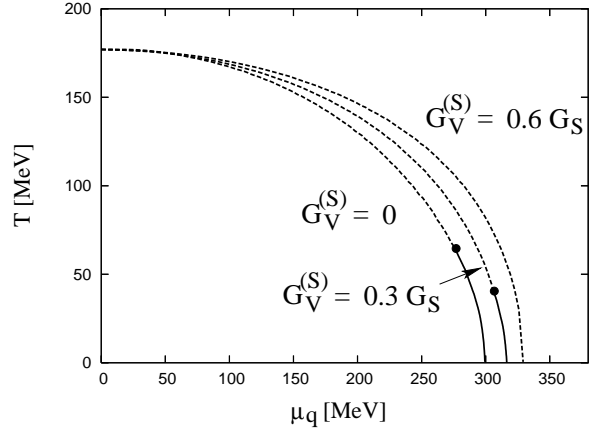}
\caption{
The phase diagram of the NJL model for an isosymmetric system
in the chiral limit~\cite{our:njl}. The tricritical point is
indicated by a dot.
}
\label{phase_eq}
\end{center}
\end{figure}

In our study of fluctuations we adopt the Nambu--Jona-Lasinio (NJL) model as an effective
chiral model under the mean field approximation~\cite{njl}. The model describes the
chiral phase transition where the dynamically generated quark mass $M$ acts as an order
parameter. In Fig.~\ref{phase_eq} we show the phase diagram of the two-flavored NJL model
for an isosymmetric system in the chiral limit. The position of the phase boundary and
the tricritical point (TCP) depends crucially  on the model parameters, like e.g. on
the strength of the four-fermion interactions. In the figure we
illustrate the dependence on the scalar-isoscalar $G_S$ and vector-isoscalar $G_V$
couplings. With increasing $G_V$, the phase transition line at fixed $T$ is shifted to
larger $\mu_q$ due to strong repulsive forces among the constituent quarks. 
Consequently,   for sufficiently large value of the vector coupling  the TCP  disappears
from the phase diagram~\cite{vector,our:njl}.

The position of the phase boundary and the order of the chiral
phase transition can be also identified through thermodynamic
observables, like net baryon number fluctuations which are sensitive
probes of the phase transition~\cite{qsus:lattice,lattice:o6,kunihiro,%
srs,qsus:model}. Furthermore, fluctuations of conserved charges are directly accessible
in experiments. Thus, it is of  importance to explore the behavior of such
fluctuations in the vicinity of the phase boundary.

The quark number and iso-vector susceptibilities, $\chi_q$ and $\chi_I$ 
respectively, describe the response of the net quark density $n_q$ and the isovector
density $n_I$ to the change of the corresponding chemical potentials. Thus, $\chi_q$ and
$\chi_I$ are defined as derivatives of $n_q$ and $n_I$  with respect to $\mu_q$ and
$\mu_I$;
\begin{equation}
\chi_q = \frac{\partial n_q}{\partial\mu_q}\,, \qquad \chi_I = \frac{\partial
n_I}{\partial\mu_I}\,,
\end{equation}
with $\mu_q$ and $\mu_I$ being the net quark and isovector chemical potential
respectively.

 The temperature dependence of $\chi_q$ shows characteristic features,
which vary rapidly with $\mu_q$. The phase boundary is signaled by a discontinuity in the
susceptibility. The size of the discontinuity grows with increasing $\mu_q$ up to the
TCP, where the susceptibility diverges. Beyond the TCP the discontinuity is again finite.
On the other hand, at $\mu_q=0$ the discontinuity vanishes and the susceptibility shows a
weaker non-analytic structure at the transition temperature, resulting in  a
discontinuity of $\partial \chi_q/\partial T$. These critical properties of
$\chi_q$ are consistent with that expected for a second order phase transition
belonging to the universality class of $O(4)$ spin model in three dimensions
\cite{srs,qsus:model,ker}.

\begin{figure}
\begin{center}
\includegraphics[width=7cm]{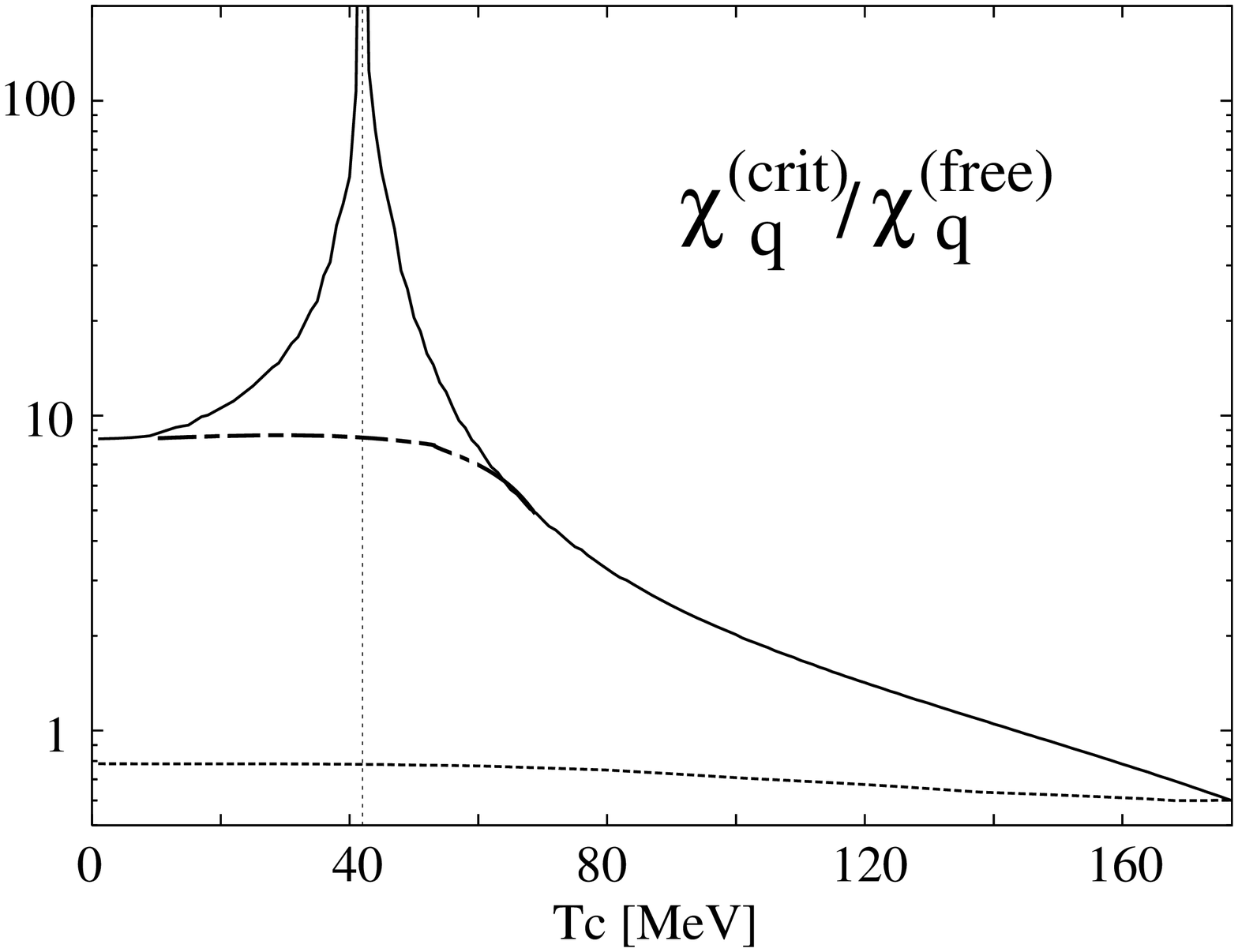}
\includegraphics[width=7cm]{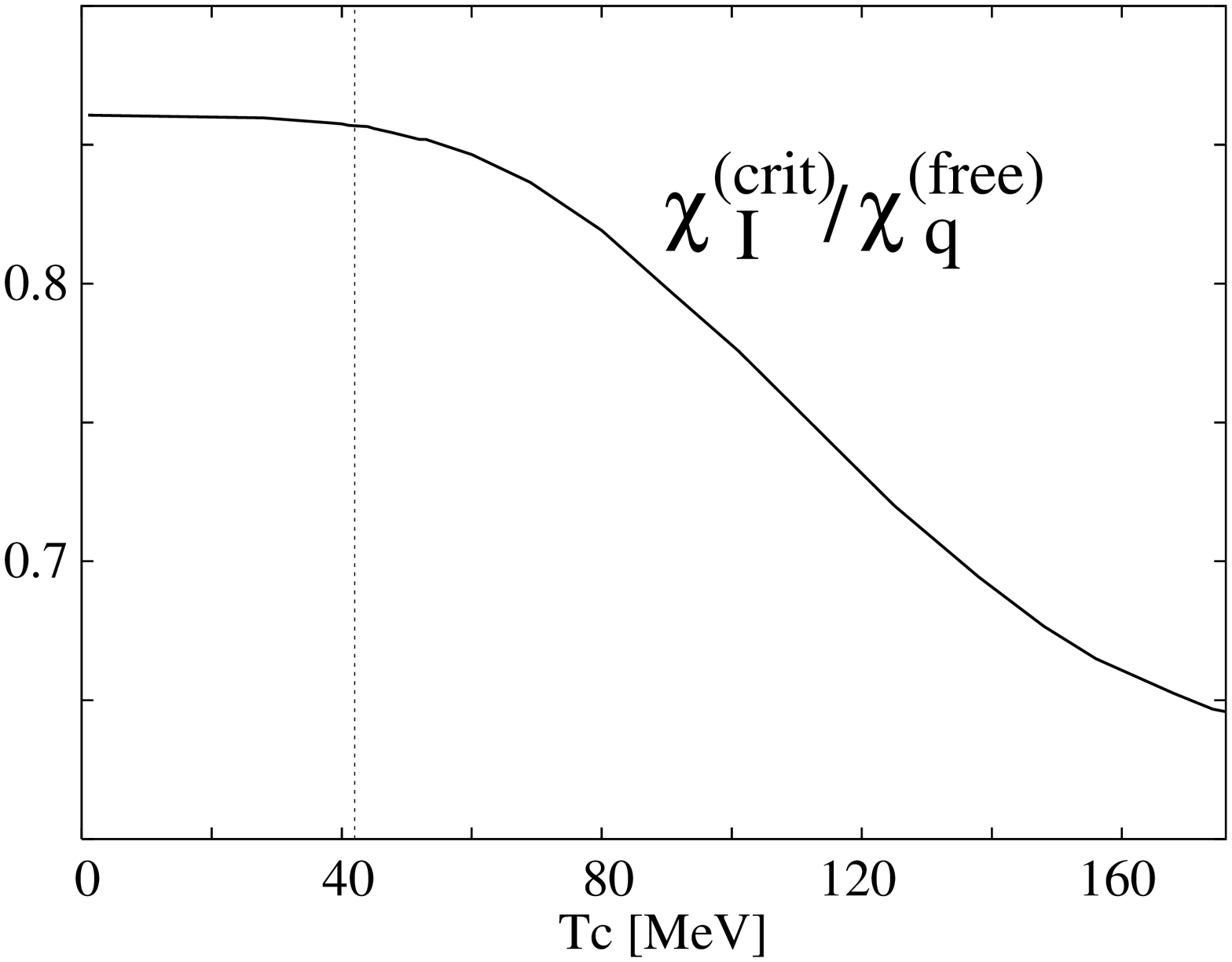}
\caption{
(Left) The net quark number susceptibility $\chi_q$ along
the phase boundary in the chiral limit. The solid (dashed) line
denotes $\chi_q$ in the chirally broken (symmetric) phase.
The vertical dotted line indicates the position of the TCP.
(Right) The isovector susceptibility along the phase
boundary~\cite{our:njl}.
}
\label{sus_cl}
\end{center}
\end{figure}

In Fig.~\ref{sus_cl} we show the net quark and isovector susceptibilities along the phase
boundary. The singularity of $\chi_q$ indicates the existence of the TCP. In the absence
of a TCP, the net quark susceptibility would be a monotonic function of $T$ along the
phase boundary, as illustrated in the Fig.~\ref{sus_cl}-left by a dashed-dotted line.
We note that on the qualitative level the  critical behavior  of the net quark
number  susceptibility can be also obtained  in the Landau
theory~\cite{qsus:model,our:njl}: First, the discontinuity across the phase boundary
vanishes at $\mu_q=0$. Second, the singularity of $\chi_q$ shows up only in the chirally
broken phase, while the susceptibility in the symmetric phase is monotonous along the
phase boundary and shows no singular behavior.
The isovector fluctuations $\chi_I$, contrary to $\chi_q$, are neither singular nor
discontinuous at the chiral phase transition for finite chemical potential. We find a
rather smooth increase of $\chi_I$ with increasing $\mu_q$ along phase boundary line.
At the TCP the $\chi_I$ remains finite. The non-singular behavior of $\chi_I$ at
the TCP is consistent with the observation that there is no mixing between isovector
excitations and the isoscalar sigma field due to isospin conservation~\cite{hs}. Recent
LGT results~\cite{lattice:o6} also show a smooth change of the isovector fluctuations
around  deconfinement transition and a fairly weak dependence of $\chi_I$ on the quark
chemical potential $\mu_q$.

The net quark number $\chi_q$ and the isovector $\chi_I$ susceptibilities are related
with fluctuations of the electric charge $\chi_Q$ as
\begin{equation}
\chi_Q=\frac{1}{36}\chi_q+\frac{1}{4}\chi_I+ \frac{1}{6} {\frac{\partial^2P}{\partial
\mu_q\partial\mu_I}}\,, \label{sus_rel}
\end{equation}
where $P$ is the thermodynamic pressure. For isospin symmetric system the last
term vanishes. Hence in this case all relevant susceptibilities are linearly dependent.
Clearly, since $\chi_I$ is finite at the TCP, the electric charge fluctuations $\chi_Q$
diverge with the same critical behavior as $\chi_q$. At finite $\mu_I$ the properties of
$\chi_I$ at the chiral phase transition change. Then, since the isoscalar sigma
field mixes with the isospin density~\cite{hs}, the isovector susceptibility exhibits a
similar structure as $\chi_q$, with a singularity at the TCP.

\begin{figure}
\begin{center}
\includegraphics[width=7cm]{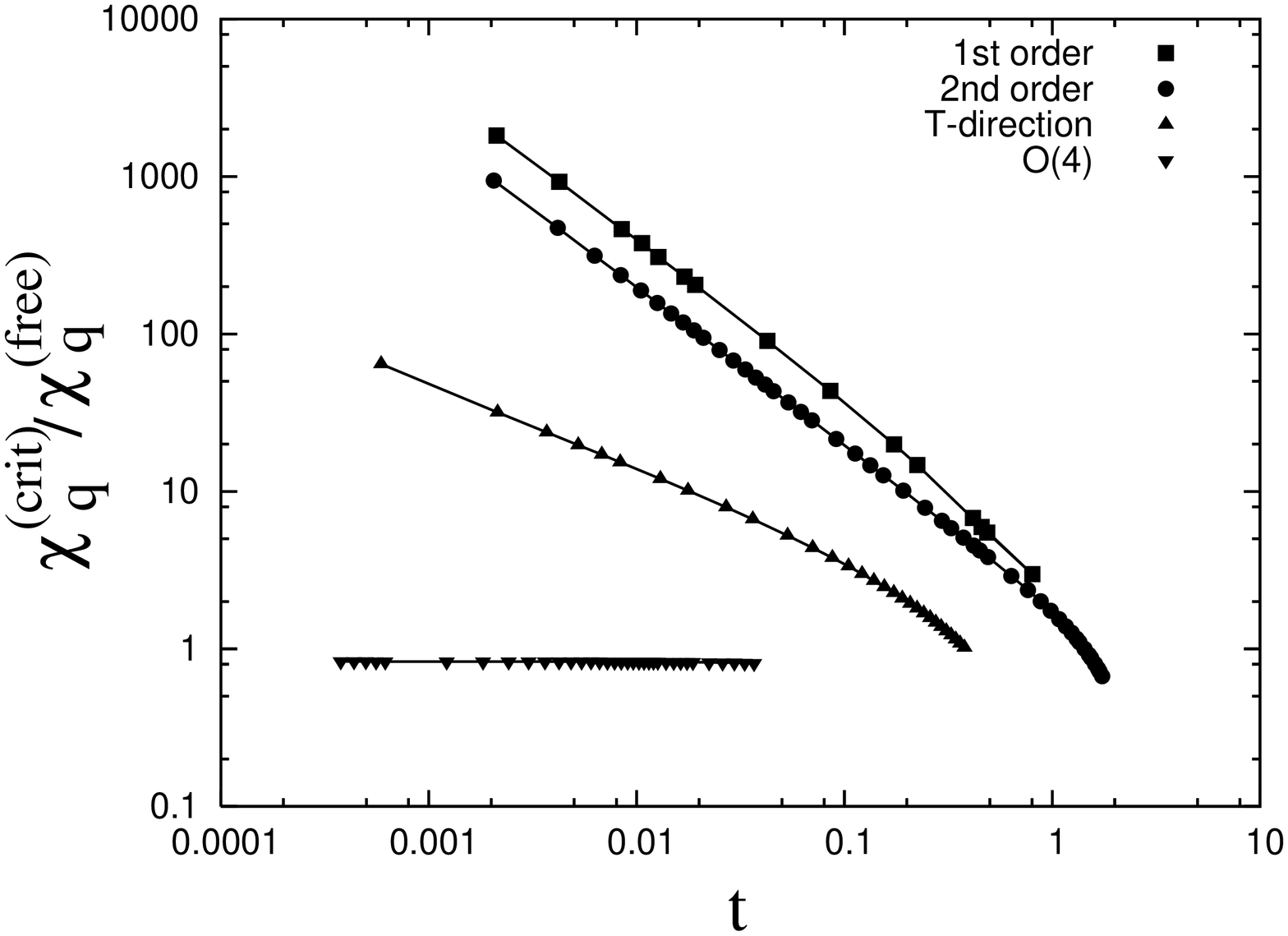}
\includegraphics[width=7cm]{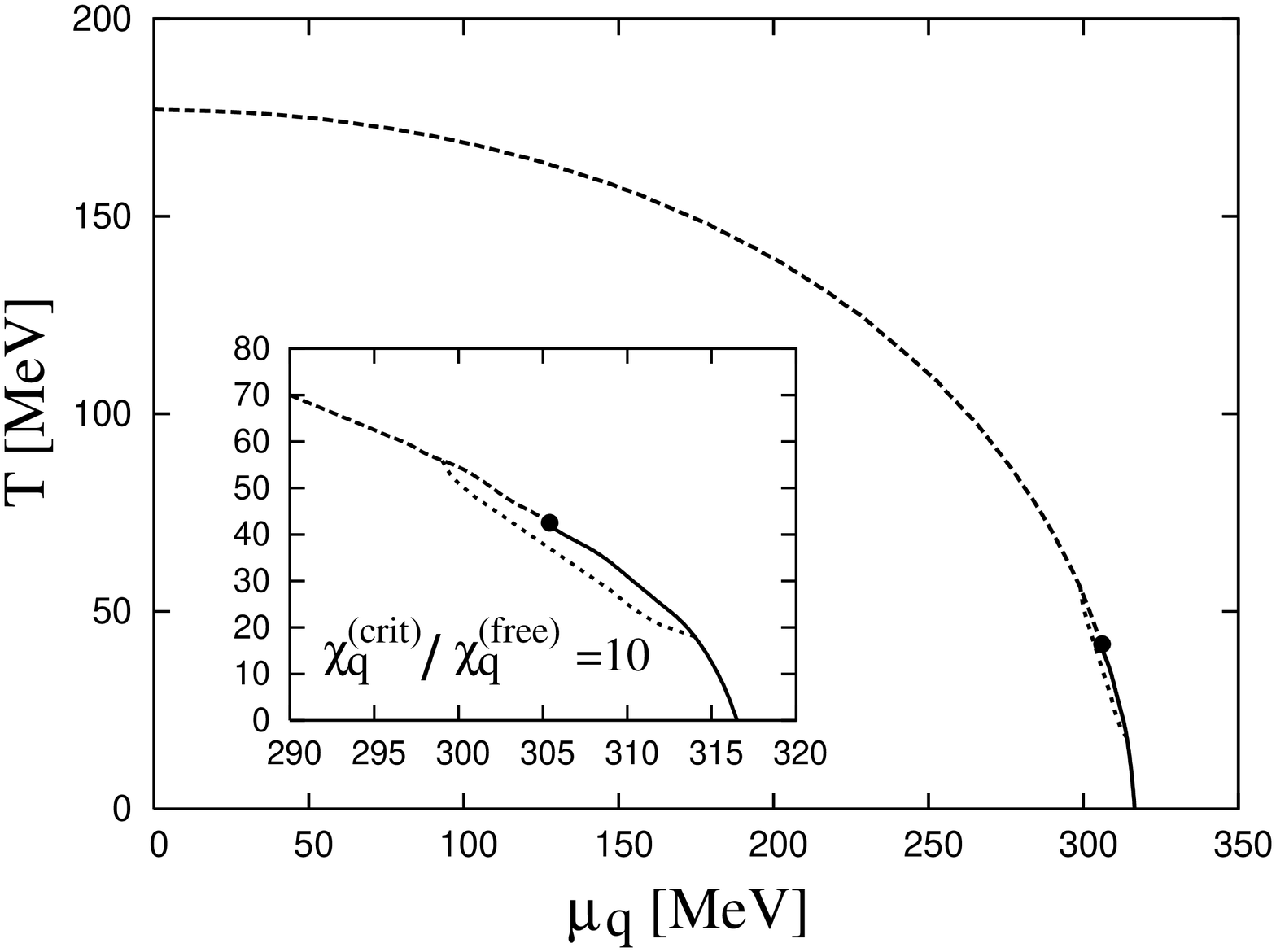}
\caption{
(Left) The quark number susceptibility near the TCP
as a function of the reduced temperature.
(Right) The critical region where the $\chi_q$ is
enhanced by an order of magnitude compared to the
free one~\cite{our:njl}.
}
\label{critical}
\end{center}
\end{figure}
The strength of singularities is governed by the critical exponents whose values are
different depending on paths approaching the phase transition \cite{gw}. The mean-field
exponents of the TCP and the CEP can be obtained from Landau theory~\cite{our:njl}. In
Fig.~\ref{critical}-left we illustrate the critical behavior of $\chi_q$ near the O(4)
critical line and at the TCP. For paths approaching a TCP asymptotically
tangential to the phase boundary, the quark number susceptibility diverges with the
critical exponent $\gamma_q=1$. On the other hand,  approaching the TCP along the
first-order transition line, the pre-factor of the singular contribution to the
quark susceptibility is twice as large as that obtained when approaching TCP  along the
O(4) critical line. For other paths the critical exponent is $\gamma_q=\frac 12$. At the
O(4) critical line, the susceptibility remains finite. The corresponding critical
exponent of the O(4) universality class is $\alpha\simeq -0.2$, while in the NJL model
under the mean-field approximation, $\alpha=0$. For non-zero quark mass, at the
critical endpoint, the mean-field critical exponent along a path not tangential to the
phase boundary is $2/3$, while along the phase boundary it remains equal to
unity~\cite{srs}. When quantum fluctuations are included, the first exponent is
renormalized to that of the 3D Ising model universality class~\cite{schaefer-wambach},
i.e. $\epsilon=0.78$.

In Fig.~\ref{critical}-right we show the ``critical'' region, where the susceptibility
exceeds its value in the ideal quark gas  by more than an order of
magnitude~\footnote{
 By ``critical'' region we mean here the region where the
 susceptibility is large due to fluctuations and  not the region
 where  the critical exponents deviates from their mean-field values.
}. The differences in values of the  critical exponents for different "paths" are
reflected in the shape of the critical region around TCP. It is elongated along the
phase boundary, where the singularity is strongest.

\section{Baryon number susceptibility in the presence of
spinodal instabilities}

In the previous section we have argued  that the enhancement  of the baryon number
fluctuations could be a clear indication for the existence of the  critical end point in
the QCD phase diagram. However, the finite  density fluctuations along the first
order transition appear under the assumption that this transition happen in
equilibrium. In non-equilibrium system,  a first order phase transition is intimately
linked with the existence of a convex anomaly in the thermodynamic pressure~\cite{ran}.
There is an interval of energy density or baryon number density where the derivative
of the pressure, $\partial P/{\partial V}>0$, is positive. This anomalous behavior
characterizes a region of instability in the ($T,n_q)$-plane which is bounded by
the spinodal lines, where the pressure derivative with respect to volume vanishes. The
derivative taken at constant temperature and that taken at constant entropy,
\begin{equation}
\left( \frac{\partial P}{\partial V} \right)_T=0 \qquad{\rm and}\qquad \left(
\frac{\partial P}{\partial V} \right)_S=0\,,
\end{equation}
define the isothermal and isentropic spinodal lines respectively.

\begin{figure}
\begin{center}
\includegraphics[width=8cm]{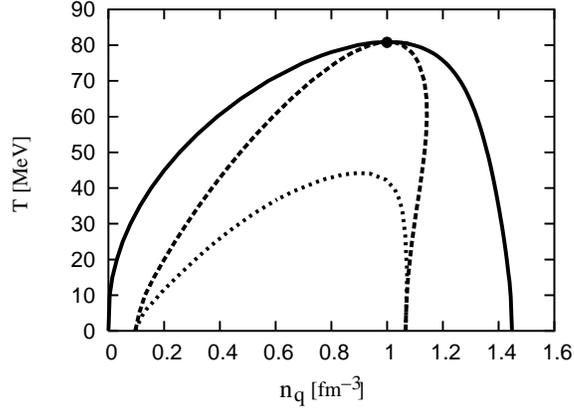}
\caption{
The phase diagram of the NJL model~\cite{our:spinodal}.
The filled point indicates the CEP. The full lines starting
at the CEP represent boundary of the coexistence region in
equilibrium. The dashed curves are the isothermal whereas the
dotted ones are the isentropic spinodal lines.
}
\label{phase_noneq}
\end{center}
\end{figure}

For finite vale of the quark mass and within large range of parameters  the NJL
model exhibits a critical end point (CEP) that separates the cross over from the first
order chiral phase transition. The relevant part of the phase diagram in the $(T,
n_q)$--plane is shown in Fig.~\ref{phase_noneq}. If the first order phase transition
takes place in equilibrium, there is a coexistence region, which ends at the CEP.
However, in a non-equilibrium first order phase transition, the system
supercools/superheats and, if driven sufficiently far from equilibrium, it becomes
unstable due to the convex anomaly in the thermodynamic pressure. In other words, in the
coexistence region there is a range of densities and temperatures, bounded by the
spinodal lines, where the spatially uniform system is mechanically unstable.

\begin{figure}
\begin{center}
\includegraphics[width=7cm]{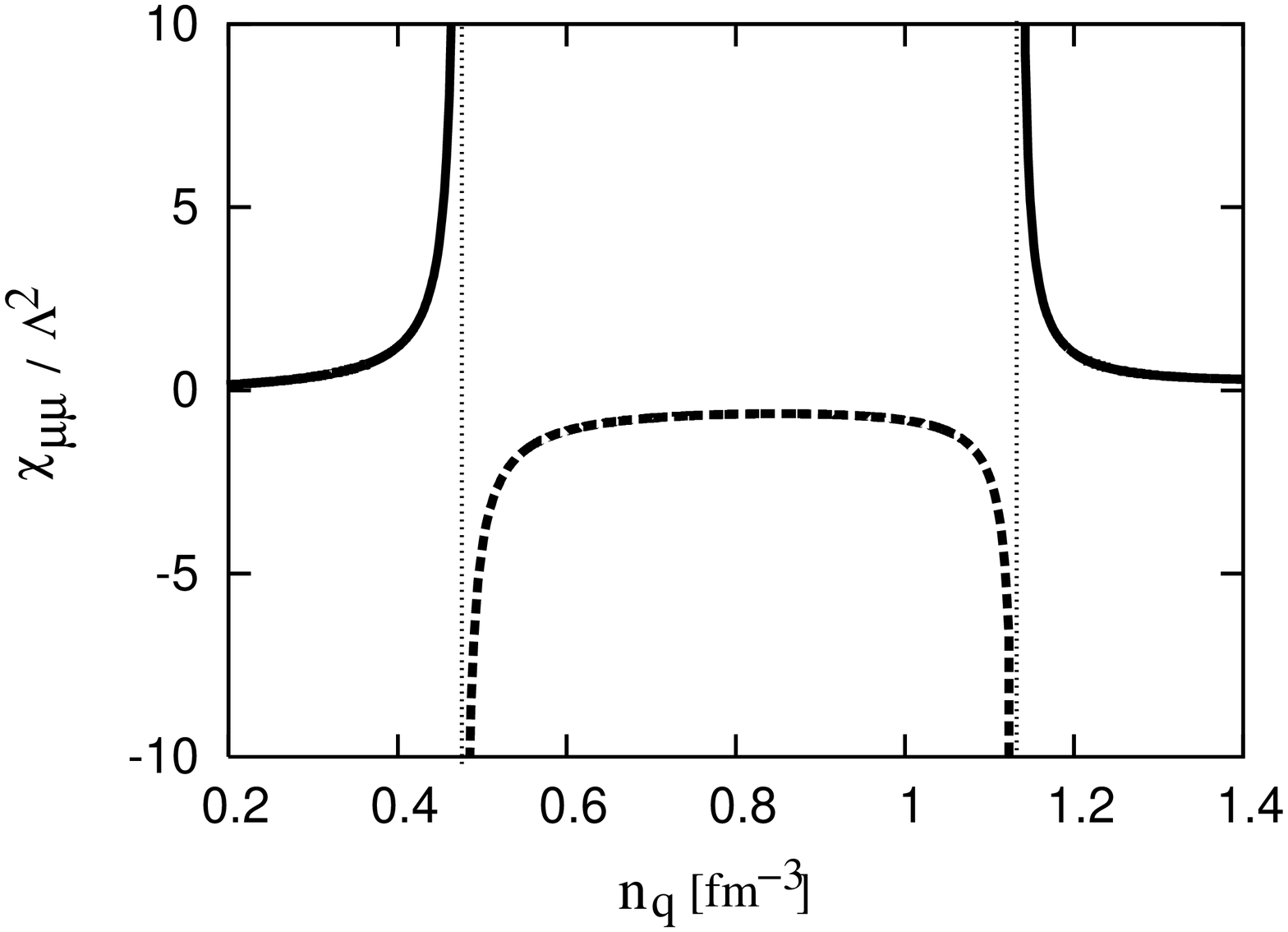}
\includegraphics[width=8cm]{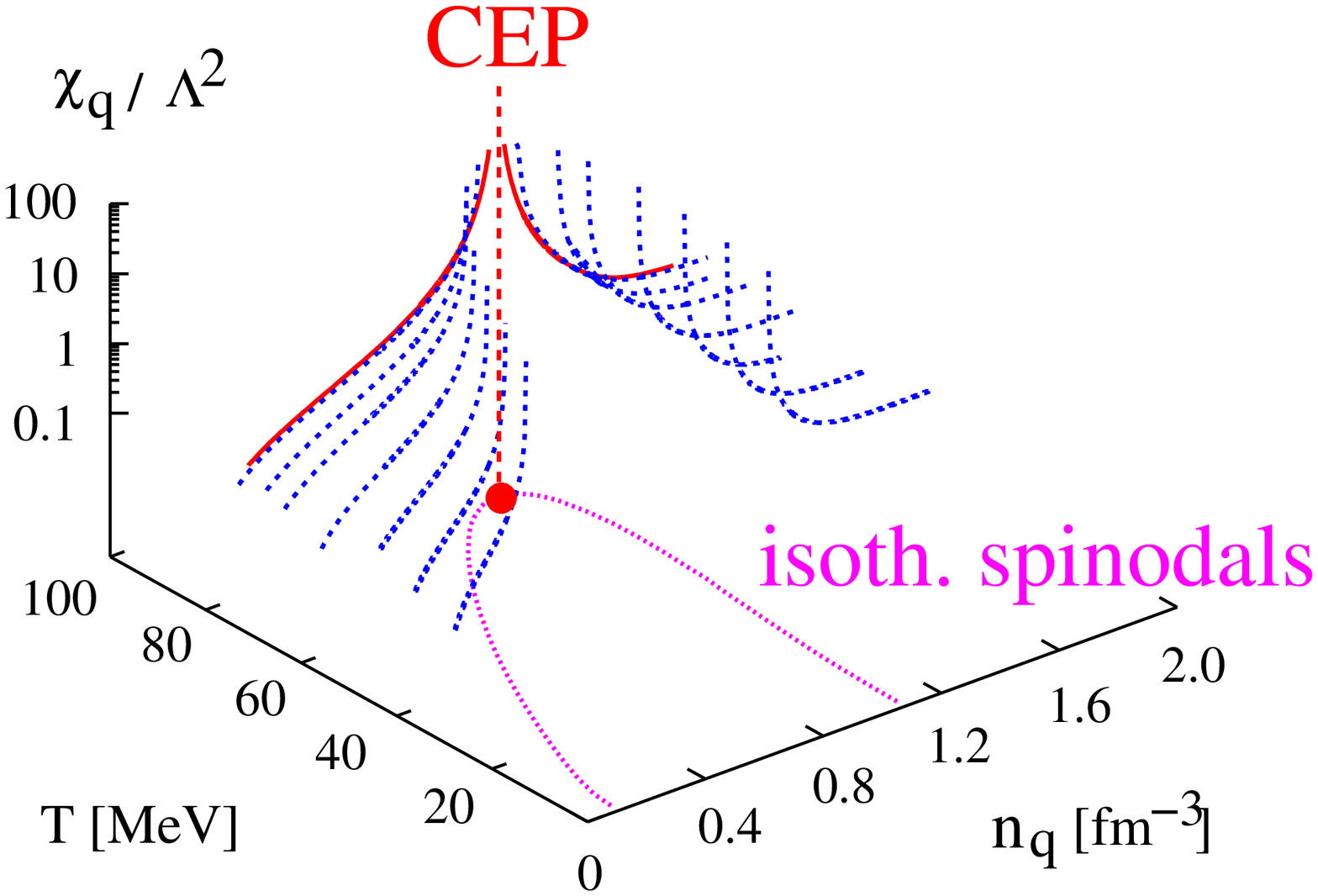}
\caption{
(Left) The net quark number susceptibility at $T=50$ MeV as a
function of the quark number density across the first order
phase transition. (Right) The net quark number susceptibility
in the stable and meta-stable regions~\cite{our:spinodal}.
}
\label{sus_sp}
\end{center}
\end{figure}

In Fig.~\ref{sus_sp}-left we show the evolution of the net quark number fluctuations
along a path of fixed $T=50$ MeV  in the $(T,n_q)$--plane. When entering the coexistence
region, there is a singularity in $\chi_q$ that appears when crossing the isothermal
spinodal lines and  where the fluctuations  changes the sign. In between the spinodal
lines, the susceptibility is negative. Consequently, this implies  instabilities in the
baryon number fluctuations when crossing from meta-stable to unstable mixed phase.
The above  behavior of $\chi_q$ is a direct consequence of the thermodynamics relation
\begin{equation}
\left( \frac{\partial P}{\partial V} \right)_T
= - \frac{n_q^2}{V}\frac{1}{\chi_q}\,.
\label{pder}
\end{equation}
Along the isothermal spinodals the  pressure derivative  in Eq. (\ref{pder}) vanishes.
Thus, for non-vanishing density $n_q$, $\chi_q$ must diverge to satisfy (\ref{pder}).
Furthermore, since the pressure derivative ${\partial P}/{\partial V}|_T$ changes sign
when crossing the spinodal line, there must be a corresponding sign change in $\chi_q$,
as seen in Fig.~\ref{sus_sp}-left. Due to the linear relation between $\chi_q$, the
isovector susceptibility $\chi_I$ and the charge susceptibility $\chi_Q$
(\ref{sus_rel}), the charge fluctuations are also divergent at the isothermal spinodals.
Thus, in heavy-ion collisions, fluctuations of the baryon number and electric charge
could show enhanced fluctuations across   the 1st order transition if   the spinodal
decomposition appears in a system.

In Fig.~\ref{sus_sp}-right we show the evolution of the  singularities   from the
spinodal lines when approaching  the CEP. The critical exponent at the isothermal
spinodal line is found to be $\gamma=1/2$, with $\chi_q \sim (\mu-\mu_c)^{-\gamma}$,
while $\gamma=2/3$ at the CEP~\cite{our:spinodal}. Thus, the singularities at the two
spinodal lines conspire to yield a somewhat stronger divergence as they join at the CEP.
The critical region of enhanced susceptibility around the TCP/CEP is fairly
small~\cite{schaefer-wambach,our:njl}, while in the more realistic non-equilibrium system
one expects fluctuations in a larger region of the phase diagram, i.e., over a broader
range of beam energies, due to the spinodal instabilities.

The rate of change in entropy with respect to temperature at constant
pressure gives the specific heat expressed as
\begin{equation}
C_P
= T \left( \frac{\partial S}{\partial T} \right)_P
= TV \left[ \chi_{TT} - \frac{2 s}{n_q}\chi_{\mu T}
{}+ \left( \frac{s}{n_q} \right)^2 \chi_q \right]\,.
\end{equation}
The entropy $\chi_{TT}$ and mixed $\chi_{\mu T}$ susceptibilities exhibit the same
behaviors as that of $\chi_q$ shown in Fig.~\ref{sus_sp}-left. Thus $C_P$ also
divergences on the isothermal spinodal lines and becomes negative in the mixed
phase~\footnote{
 The specific heat with constant volume, on the other hand,
 continuously changes with $n_q$ and has no singularities
 on the mean-field level.
}. It was  argued that in low energy nuclear collisions the negative specific
heat could be a signal of the liquid-gas phase transition~\cite{chomaz}. Its occurrence
has recently been reported as the first experimental evidence for such an anomalous
behavior ~\cite{experiment}.

\section{Conclusions}

We presented a brief discussion of probing the QCD phase structure. We have especially
discussed the importance of conserved charge fluctuations. It was shown that the net
baryon number susceptibility must yield large contribution around the critical end point
(CEP). Consequently, a non monotonic behavior of these fluctuations as functions of the
collision energy in heavy ion collisions could  be considered as an  indication of
the CEP in the QCD phase diagram.

We have also shown that in the presence of spinodal instabilities the above picture is
modified: The net quark number fluctuations diverge at the isothermal spinodal lines of
the first order chiral phase transition. As the system crosses this line, it becomes
unstable with respect to spinodal decomposition. The unstable region is in principle
reachable in non-equilibrium systems that is most likely  created in heavy ion
collisions. Consequently,  large fluctuations of baryon and electric charge  densities
are expected not only at the CEP but also when system crosses  a non-equilibrium first
order transition.

\section*{Acknowledgments}

The work of B.F. and C.S. was supported in part by the Virtual
Institute of the Helmholtz Association under the grant No. VH-VI-041.
C.S. also acknowledges partial support by DFG cluster of excellence
Origin and Structure of the Universe.
K.R. acknowledges partial support of the Gesellschaft f\"ur
Schwerionenforschung (GSI) and the Polish Ministry of National
Education (MEN).


\end{document}